\documentstyle[12pt]{article}
\def\be{\begin{equation}}
\def\ee{\end{equation}}
\def\bea{\begin{eqnarray}}
\def\eea{\end{eqnarray}}

\begin{document}

\begin{titlepage}

\title{Spherically symmetrical configurations of self-dual Yang-Mills\\
and Einstein-Plebanski equations}
\author{ A.N.Leznov$^{1,2,3}$, P.A.Marquez Aguilar$^{1}$ and S. Mansurova$^{1}$\\
$^1${\it Research Center of Engeneering and Applied Sciences}\\
{\it Av. Universidad 1001 col. Chamilpa C.P. 62210}\\
{\it Cuernavaca Morelos Mexico.}\\
$^2${\it  Institute for High Energy Physics, 142284 Protvino,}\\
{\it Moscow Region,Russia}\\
{\it and}\\
$^3${\it  Bogoliubov Laboratory of Theoretical Physics, JINR,}\\
{\it 141980 Dubna, Moscow Region, Russia.}}

\maketitle

\begin{abstract}
Spherically symmetrical reductions of self-dual Yang-Mills
and Einstein-Plebanski equations are constructed at the same manner.
As in the first case we come back to known before solutions (under such kind
of reduction but in some different form) if in the second one we obtain
unknown before equation describing the spherically symmetrical configurations
of Plebanski equation and some number of its particular solutions.
\end{abstract}

\end{titlepage}

\section{Introduction}

Outstanding similarity the form of self-dual Yang-Mills and Plebanski \cite
{PL} heanlivy equations give to many investigators possibility to assume
that these two systems (really system in the first case and single equation
in the second one) possesses the solutions of the similar kind (of course in
some sense).

In the present paper we would like to investigate this problem on the
example of spherically symmetrical solutions of such systems. In the case of
self-dual Yang-Mills system (for arbitrary semisimple guage algebra and
arbitrary $A_1$ subalgebra embedding in it) such kind of solution were
obtained 20 years ago in \cite{LS}. Corresponding result with respect to
Plebanski equation to the best of our knowledge is absent up to now.

We will use the following forms (in each case one of many other possible
ones) for Y-M and P equations
\begin{equation}
M_{y,\bar y}+M_{z,\bar z}=[M_y,M_z],\quad P_{y,\bar y}+P_{z,\bar z}=
P_{y,y}P_{z,z}-P^2_{y,z}\equiv\{P_y,P_z\}_{y,z}  \label{MP}
\end{equation}
where function $M$ takes values in arbitrary semisimple algebra, $P$ is
scalar function.

We will use the same reducing procedure to spherically symmetrical case to
both equations (\ref{MP}), which allow as to compare and emphasize essential
differences in either cases.

\section{Yang-Mills case}

Let us use for solution of (\ref{MP}) the following ansatz
\begin{equation}
M=(\bar y)^{-1}m(r,t),\quad r^2=y\bar y+({\frac{z+\bar z}2})^2,\quad t={%
\frac{z-\bar z}{2i}}  \label{AYM}
\end{equation}
For derivatives involved in (\ref{MP}) we obtain
\[
M_y={\frac 12}{\frac{m_r}r},\quad M_{y,\bar y}={\frac y4}({\frac{m_r}r})_r
\]
\[
M_z=(\bar y)^{-1}({\frac{z+\bar z}4}{\frac{m_r}r}+{\frac 1{2i}}m_t),\quad
M_{zz}=(\bar y)^{-1}(({\frac{z+\bar z}4})^2({\frac{m_r}r})_r+{\frac 14}%
m_{tt})
\]
Substituting these expressions into first equation (\ref{MP}) we obtain
\begin{equation}
m_{rr}+m_{tt}={\frac 1{2ir}}[m_t,m_r]  \label{SS}
\end{equation}
After introduction of two complex conjugative variables $\xi =t+ix,\bar \xi
=t-ix$ we are able to rewrite (\ref{SS}) in the form of the system of
equations of the main chiral field with the moving poles
\begin{equation}
(\xi -\bar \xi )m_{\xi ,\bar \xi }=[m_\xi ,m_{\bar \xi }]  \label{EE}
\end{equation}
The technique of integration of (\ref{EE}) reader can find in \cite{ANL}.

\section{Plebanski equation}

In this case the ansatz (\ref{AYM}) is slightly modificates \footnote{%
The different degreeses on $\bar y$ before the invariant function reflects
the facts of different spins 1 and 2 in both cases}
\begin{equation}
P=(\bar y)^{-2}p(x,t),\quad x\equiv r^2=y\bar y+({\frac{z+\bar z}2})^2,\quad
t={\frac{z-\bar z}{2i}}  \label{AYM}
\end{equation}

For the derivatives involved in Plebanski equation we have in a consequent
\[
P_y=(\bar y)^{-1}p_x,\quad P_{yy}=p_{xx},\quad P_{yz}=(\bar y)^{-1}(p_{xx}{%
\frac{z+\bar z}2}+p_{xt}{\frac 1{2i}}),
\]
\[
P_z=(\bar y)^{-2}(p_x{\frac{z+\bar z}2}+p_t{\frac 1{2i}}),\quad P_{zz}=(\bar
y)^{-2}(p_{xx}({\frac{z+\bar z}2})^2+p_{xt}{\frac{z+\bar z}{2i}}+{\frac 12}%
p_x-{\frac 14}p_{tt}),
\]
\[
P_{y,\bar y}=(\bar y)^{-2}(y\bar yp_{xx}-p_x),\quad P_{z,\bar z}=(\bar
y)^{-2}(p_{xx}({\frac{z+\bar z}2})^2+p_{tt}{\frac 14}+{\frac 12}p_x)
\]
Substituting these expressions into Plebanski equation (\ref{MP}) we obtain
\[
p_{xx}p_{tt}-p_{xt}^2=2p_xp_{xx}+2p_x-p_{tt}-4xp_{xx}
\]
and after exchanging of unknown function $p\to p-{\frac{x^2}2}$ we rewrite
the previous equation in the form
\begin{equation}
p_{xx}p_{tt}-p_{xt}^2=2p_xp_{xx}-6xp_{xx}+4x=(p_x^2-6(xp_x-p)+2x^2)_x\equiv
F_x  \label{PF}
\end{equation}
The last is nonhomogeneous Monge-Ampher equation. The regular methods for
its integration are unknown.

Let us formally rewrite (\ref{PF}) in form of Poisson breackets
\[
(p_x)_F(p_t)_t-(p_x)_t(p_t)_F=1
\]
and resolve the last equation with the help of generating function of
canonical transformation $W(p_x,t)$
\begin{equation}
p_t=W_{p_x}(p_x,t),\quad F=p_x^2-6(xp_x-p)+2x^2=W_t(p_x,t)  \label{W}
\end{equation}
But under such kind of the procedure generating function $W$ is not
arbitrary one but satisfy some equation obtaining of which is our nearest
goal. Let us differentiate two equations (\ref{W}) with respect to x and t.
Result of differentiation of the first equation is the following
\begin{equation}
p_{tx}=W_{p_x,p_x}p_{xx},\quad p_{tt}=W_{p_x,p_x}p_{tx}+W_{p_x,t}  \label{A}
\end{equation}
and the same with respect to second equation (\ref{W})
\begin{equation}
2p_xp_{xx}-6xp_{xx}+4x=W_{p_x,t}p_{xx}\quad
2p_xp_{xt}-6xp_{tx}+6p_t=W_{p_x,t}p_{xt}+W_{tt}  \label{B}
\end{equation}
Multiplying the second equation (\ref{A}) on $p_{xx}$ and keeping in mind
two first equations (\ref{A}) and (\ref{B}) we conclude that equation (\ref
{PF}) is satisfied. Resolving (\ref{B}) with respect to $p_{xx},p_{xt}$ we
obtain ($p_t=W_{p_x}$)
\begin{equation}
p_{xx}={\frac{4x}{W_{t,p_x}+6x-2p_x}},\quad p_{xt}={\frac{6W_{p_x}-W_{tt}}{%
W_{t,p_x}+6x-2p_x}}  \label{C}
\end{equation}
Combining (\ref{C}) with the first equation (\ref{A}) we able to express $x$
in term of derivatives of the generating function
\begin{equation}
4x={\frac{6W_{p_x}-W_{tt}}{W_{p_x,p_x}}}  \label{D}
\end{equation}
Result of differentiation of the last equation with respect to $x,t$
arguments lead to
\begin{equation}
4=({\frac{6W_{p_x}-W_{tt}}{W_{t,p_x}}})_{p_x}p_{xx},\quad 0=({\frac{%
6W_{p_x}-W_{tt}}{W_{t,p_x}}})_{p_x}p_{xt}+({\frac{6W_{p_x}-W_{tt}}{W_{t,p_x}}%
})_t  \label{E}
\end{equation}
Keeping in mind that $p_{xx},p_{xt}$ as a corollary of (\ref{C}) and (\ref{D}%
) are expressed in terms of the derivatives of generating function $W$, we
conclude that (\ref{E}) really is a system of equations which this function
must satisfy. Introducing the notation $\alpha ={\frac{6W_{p_x}-W_{tt}}{%
W_{p_x,p_x}}}$ we rewrite (\ref{E}) as
\[
\alpha _t+4W_{p_x,p_x}=0,\quad \alpha \alpha _{p_x}=4W_{p_x,t}+6\alpha -8p_x
\]
Let us substitute $\alpha =\beta _{p_x}$. The last substitution allows to
integrate once the previous system with the result
\begin{equation}
\beta _t=-4W_{p_x}+A(t),\quad {\frac{\beta _{p_x}^2}2}=4W_t+6\beta
-4p_x^2+B(t)  \label{G}
\end{equation}
From definition of $\alpha $ function
\[
\alpha =\beta _{p_x}={\frac{6W_{p_x}-W_{tt}}{W_{p_x,p_x}}}
\]
and known expressions for derivatives of $W$ function via $\beta $ function (%
\ref{G}) it follows immediately that
\[
B_t(t)+A(t)=0
\]
And at last after excluding $W$ from system (\ref{G}) we come to a single
equation which function $\beta $ satisfy
\begin{equation}
\beta _{tt}+({\frac{\beta _{p_x}^2}2})_{p_x}=-8p_x+6\beta _{p_x}+A_t
\label{BBB}
\end{equation}
After differentiation of the last equation once with respect to $p_x\equiv y$
we come to equation, which $\alpha $ function satisfy
\begin{equation}
\alpha _{tt}+({\frac{\alpha ^2}2})_{yy}=-8+6\alpha _y  \label{BBB1}
\end{equation}

\section{Some particular solutions}

Here we would like to present some number of particular solutions, arising
as a result of the corresponding reductions. We like to emphasize here that
all equations (and systems) below allow the general solution.

\subsection{Anzats $\alpha=x^2 A(t)+x B(t)+C(t)$}

Substituting the anzats of the title of this subsection into (\ref{BBB1})
and equating to zero coefficient on $0,1,2$ degrees of $x$ variable we
obtain
\[
x^2\ddot A(t)+x\ddot B(t)+\ddot
C(t)=2A(x^2A(t)+xB(t)+C(t))+(2Ax+B)^2=-8+6(2Ax+B)
\]
\begin{equation}
\ddot A+6A^2=0,\quad \ddot B+6AB=-12A,\quad \ddot C+2AC=-(B-2)(B-4)
\label{B}
\end{equation}
Equation for $A$ function can be resolved in terms of particular case of
Weirstrasse function \cite{WV}
\[
\dot A=\sqrt {C-4A^3}.
\]
Equation for $B$ is integrable in quadratures with the result
\[
B=-2+c_1A+c_2\int {dt\over B^2}
\]
In terms of the Weirstrasse like function  may be resolved in quadratures
equation for $C$ function (see \cite{WV}).

\subsection{The travelling wave $\alpha\equiv \alpha(x+vt)$}

Under such substitution the ordinary differential equation takes the form
\begin{equation}
(\alpha+v^2)\alpha^{\prime\prime}+(\alpha^{\prime})^2=-8+6\alpha^{\prime}
\label{RV}
\end{equation}
where $\alpha^{\prime}=\alpha_{\xi},\xi=x+vt$. After introducing $\tilde
\alpha= \alpha+v^2$ the last equation allows integration ones with the
result
\[
\tilde \alpha \tilde \alpha^{\prime}=-8(\xi+c_1)+6\tilde \alpha
\]
Consequent exchangings $\xi\to \xi+c_1, \tilde \alpha=\xi \nu$ leads the
last equation to a form
\[
\xi \nu^{\prime}+\nu==-{\frac{8}{\nu}}+6
\]
The way of integration of the last equation is obvious with the result
\begin{equation}
{\frac{\nu d\nu}{(\nu-2)(\nu-4)}}=-d\ln \xi,\quad {\frac{(\nu-4)^2}{\nu-2}}%
\xi=c_2  \label{NUS}
\end{equation}
Returning to initial variables and functions we obtain the following general
solution of (\ref{RV}) (but only particular solution of (\ref{BBB1}))
\begin{equation}
\alpha=-v^2+{\frac{(8(x+vt)+\tilde c_1)\pm \sqrt {c_2(8(x+vt)+\tilde c_1})}{2%
}}, \quad \tilde c_1=c_2+8c_1  \label{FFF}
\end{equation}

\subsection{Automodel solution}

Equation (\ref{BBB1}) is obviously invariant with respect to transformation
\[
\alpha (t,x)=p^{-1}\alpha (\sqrt{p}t,px)
\]
As a corollary it arised automodel solution of this equation
\[
\alpha =xf({\frac x{t^2}})
\]
Using anzats above we obtain for corresponding derivatives $(\xi \equiv {%
\frac{x\ }{t^2}})$
\[
\alpha _x=(\xi f)_\xi ,\quad \alpha \alpha _{xx}=(\xi f)(\xi f)_{\xi ,\xi }
\]
\[
\alpha _t=-2t\xi ^2f_\xi ,\quad \alpha _{tt}=-2\xi ^2f_\xi +4\xi (\xi
^2f_\xi )_\xi
\]
Introducing the new unknown function $\nu \equiv \xi f$ we rewrite (\ref
{BBB1}) $(\nu _\xi \equiv \nu ^{\prime })$
\begin{equation}
-2(\xi \nu ^{\prime }-\nu )+4\xi ^2\nu ^{\prime \prime }+\nu \nu ^{\prime
\prime }=-(\nu ^{\prime }-2)(\nu ^{\prime }-4)  \label{NU}
\end{equation}
We present below three its different particular solutions, which can be
checked by direct not cumbersome calculations
\begin{equation}
\nu =-\xi ^2+2\xi +c,\quad \nu =c\xi -{\frac{(c-2)(c-4)}2},\quad \nu =4\xi
+c\xi ^{{\frac 12}}  \label{PS}
\end{equation}
All these solutions are depending on only one arbitrary parameter $c$, while
the general solution of (\ref{NU}) must contain two arbitrary parameters.

To obtain general solution it is necessary to consider not obvious and
simple but very important symmetry of (\ref{NU}).

With this goal let us perform two consequent transformations
\[
\xi\to \xi+q,\quad \nu\to \tilde \nu-8q \xi-4q^2
\]
As a result of such transformations equation (\ref{NU}) takes the form
\begin{equation}
-2(\xi \tilde \nu^{\prime}-\tilde \nu)+4\xi^2 \tilde
\nu^{\prime\prime}+\tilde \nu \tilde \nu^{\prime\prime}= -(\tilde
\nu^{\prime})^2+(6+18q)\tilde \nu^{\prime}-(2+8q)(4+8q)-8q^2  \label{NUM}
\end{equation}
Choosing $q=-{\frac{1}{3}}$ we obtain finally
\begin{equation}
-2(\xi \tilde \nu^{\prime}-\tilde \nu)+4\xi^2 \tilde
\nu^{\prime\prime}+\tilde \nu \tilde \nu^{\prime\prime}= -(\tilde
\nu^{\prime})^2  \label{NUF}
\end{equation}

The last equation is obviously invariant with respect to transformation
\[
\tilde \nu(\xi)\to p^{-2} \tilde \nu(p\xi)
\]
Gathering all results together we obtain in consequent. If $\nu(\xi)$ is
solution of (\ref{NU}) than both $\tilde \nu$ are solutions of the equation (%
\ref{NUF}).
\[
\tilde \nu=\nu(\xi-{\frac{1}{3}})-{\frac{8\xi}{3}}+{\frac{4}{9}},\quad
\tilde \nu=p^{-2}(\nu(p\xi-{\frac{1}{3}})-{\frac{8p\xi}{3}}+{\frac{4}{9}})
\]
Performing the inverse transformation we return to a new solution of the
initial system in the form
\begin{equation}
\nu=p^{-2}[\nu(p(\xi+{\frac{1}{3}})-{\frac{1}{3}})-{\frac{8p(\xi+{\frac{1}{3}%
})}{3}}+ {\frac{4}{9}}]+{\frac{8(\xi+{\frac{1}{3}})}{3}}-{\frac{4}{9}}
\label{FF}
\end{equation}

(\ref{FF}) is the mentioned above not simple and obvious invariant
transformation of the equation (\ref{NU}).

Two first solutions of (\ref{PS}) are covariant with respect to such
transformation -- solution preserve their forms with changed value of the
constant of integration. In the first case $c\to p^{-2}(c-{\frac{p^2}{9}}- {%
\frac{1}{3}})+{\frac{4}{9}}$ in the second one $c\to p^{-1} c-p^{-1}{\frac{8%
}{3}}+ {\frac{8}{3}}$.

But third solutions after such transformation does not preserve its form and
passes to a general solution of the system (\ref{NU}) dependent on two
arbitrary constants $p$ and $c$
\begin{equation}
\nu =p^{-2}[4(p(\xi +{\frac 13})-{\frac 13})+c\sqrt{p(\xi +{\frac 13})-{%
\frac 13}}-{\frac{8p(\xi +{\frac 13})}3}+{\frac 49}]+{\frac{8(\xi +{\frac 13}%
)}3}-{\frac 49}  \label{FFF}
\end{equation}
That (\ref{FFF}) is really (general) solution of (\ref{NU}) one can
convinced by direct computation. Going back to $\alpha $ we obtain the
particular solution of the initial equation (\ref{BBB1})
\begin{equation}
\alpha =({\frac 43}p^{-1}+{\frac 89})x+{\frac 49}p^{-2}(p-1)(p+2)t^2+p^{-2}ct%
\sqrt{px+{\frac{p-1}3}t^2}  \label{F34}
\end{equation}

\section{Outlook}

The main result of the present paper consists in new equation (\ref{BBB1})
describing the spherically symmetrical configurations of Plebanski equation
and some number of its particular solutions. At this moment we are unable to
say anything definite is this equation integrable or not. At least to the list
of known up to now integrable equations it doesn't belongs. We hope in future
publications find the answer on this intrigued question.

\end{document}